\nonstopmode{}
\documentclass[aps,twocolumn,showpacs,pre,superscriptaddress]{revtex4}
\usepackage{graphicx,color}
\newcommand{\figwidth}{3.375in}
\newcommand{\figwidtha}{3.2in}

\begin{document}
\title{Thermodynamic Construction of an One-Step Replica-Symmetry-Breaking
Solution in Finite Connectivity Spin Glasses} 
\author{T. Nakajima}
\email{tetsuya@huku.c.u-tokyo.ac.jp}
\affiliation{Department of Basic Science, the University of Tokyo, 3-8-1 Komaba, Meguro-ku,Tokyo 153-8902 Japan}
\author{K. Hukushima}
\email{hukusima@phys.c.u-tokyo.ac.jp}
\affiliation{Department of Basic Science, the University of Tokyo, 3-8-1
Komaba, Meguro-ku,Tokyo 153-8902 Japan}
\date{\today}
\begin{abstract}
An one-step replica-symmetry-breaking solution for finite connectivity 
 spin-glass models with $K$ body interaction is constructed at finite
 temperature using the replica method and thermodynamic constraints. 
In the absence of external fields, this construction provides a general
 extension of replica symmetric 
 solution at finite replica number to one-step replica-symmetry-breaking
solution. It is found that this result is formally equivalent to that
 of the one-step replica-symmetry-breaking cavity method.
 To confirm the validity of the obtained solution, Monte Carlo
 simulations are performed for $K=2$ and $3$.
The thermodynamic quantities of the Monte Carlo results extrapolated to
 a large-size limit are consistent with 
those estimated by our solution for $K=2$ at all  simulated temperatures
 and for $K=3$ except near the transition temperature. 
 \end{abstract}
\maketitle
\section{Introduction}

Since the celebrated paper by Edwards and Anderson\cite{EA1975}, 
mean-field theory of spin glass (SG) has been extensively investigated.
The replica theory\cite{P1980,MPV1987} is one of the most successful
achievement that  
has revealed the nature of the low temperature phase of mean-field SG
models. Parisi's pioneering work provided the replica method with
implementation of replica symmetry breaking (RSB).
Originally, $K$ step RSB ($K$RSB) was proposed as ``a sequence of 
approximated solutions'' to the true solution and the full RSB solution
was derived as a $K\to\infty$ limit. This approach has actually proven to be
exact recently\cite{T2006} for the Sherrington-Kirkpatrick (SK)
model\cite{SK1978}.   
Although this introduction of RSB is motivated
by de Almeida-Thouless (AT) condition\cite{AT1978}, which is the instability
of replica symmetric (RS) solution with respect to replica
couplings, it should be noted that AT instability is only one of the
possible scenario for RSB\cite{K1983} and that 
the origin of RSB is in general model-dependent.
In addition,
a 1RSB solution for various mean-field SG models\cite{G1985,GKS1985} is stable
with respect to further RSB perturbation, and $K$RSB rarely appears for
$K\geq 2$.
These facts suggest that 
there is another mechanism to break the replica symmetry and 
it distinguishes 1RSB from full RSB (FRSB). 

Recently, the authors have shown\cite{NH2008} that $p$-body SK model, which is a
typical model to exhibit a SG transition to 1RSB phase,
actually has another reason to break the replica symmetry above the Gardner temperature\cite{G1985}.
It is the monotonicity condition of the cumulant generating function of 
the free energy $\phi(n)$, whose limiting value at $n=0$ is the
averaged free energy,  rather than the AT condition
that causes RSB\cite{NH2008}.
The relevance of these conditions is reversed at the Gardner
temperature, where the transition between 1RSB and full RSB takes place.
Furthermore, it is proved that if the monotonicity is broken 
in the absence of external field, which ensures the smallest overlap parameter
$q_0=0$, then the correct 1RSB solution is given by the RS solution at $n_m$,
which is defined as the monotonicity breaking point, $i.e.$,
$\phi'(n_m)=0$. 
This has revealed that the continuation of the cumulant
generating function $\phi(n)$ to $\phi(0)$ is strongly restricted 
by a kind of thermodynamic constraints and that it naturally induces the
1RSB solution in the case of a fully connected mean-field SG model. 
Regarding $n$ as a fictitious inverse temperature, we can resort to
the thermodynamics for extracting high-temperature, or replica,
limit($n\to 0$) from low-temperature behavior($n\gg 1$). 
These facts strongly suggest  
that 1RSB is a consequence of the monotonicity breaking and FRSB 
is that of AT stability breaking. 

Finite connectivity SG model has been considered as a first non-trivial
extension of the mean-field theory, and challenged in many literatures. 
As a straight-forward extension from the case of fully connected model,
perturbation theories in the region of the large connectivity or
near the transition temperature have been studied in the replica
formalism\cite{LG1989,GD1990}.  
Another replica calculation\cite{MP2001,FLRZ2001,FMRWZ2001} has succeeded to 
derive an exact expression of the free energy under a non-trivial
ansatz called factorized ansatz. 
The difficulty in these works appears in the search for an RSB
saddle-point, because RSB is defined using the symmetry of a
saddle-point in the theory.
In contrast, the cavity method turned out to be an alternative and
promising approach to study the finite connectivity models within 1RSB
scheme\cite{MP2001,MR2004,FLT2003,MRZ2003,KZ2008}. The key concept of this
method is the complexity\cite{M1995}, logarithm of a number of the pure
states, which enables one to deeply understand the microscopic structure
of configuration space. It is found that  the non-negativity condition
of the complexity is relevant for the 1RSB cavity scheme, that provides
a general procedure for mean-field type models including finite
connectivity SG.

In this paper, 
we further examine the possibility of 1RSB 
scenario suggested in our previous work, which might be important for 
a better understanding of the SG theory and also the replica method
itself. 
The model discussed is a finite-connectivity Ising SG model with
$K$-body interactions. The reason why this model is considered as a
good example is twofold. First our construction of
1RSB solution is applicable to the finite-connectivity SG model,
because RS solution can be explicitly obtained. 
Second, we see a direct correspondence between 
the guiding principle of introducing 1RSB in the 
replica method and the cavity method\cite{MP2001}. 

The organization of this paper is as follows. In Sec.~\ref{sec:rep},
we review our previous work\cite{NH2008} 
for complete and detailed instructions of our scheme,
in which a construction
of a 1RSB solution from RS ansatz is explained. Then a SG
model defined on 
a sparse random graph is introduced and the 1RSB solution for
the model obtained by our scheme is presented. 
We also discuss a relationship between our scheme based
on the replica theory and the cavity method for the model.
In Sec.\ref{sec:num}, 
we compare the 1RSB solution to the result by MC simulation. 
Finally Sec.~\ref{sec:sum}
is devoted to our conclusions and discussions.

\section{Model and Replica Analysis}
\label{sec:rep}
\subsection{Preliminary}
\label{sec21}
In this section, we briefly review our previous work\cite{NH2008} and
explain our scheme for the construction of a 1RSB solution in a general
manner. 
For a given Hamiltonian $H$, equilibrium statistical mechanics requires
to calculate 
the partition function $Z = {\rm Tr}\exp(-\beta H)$, where 
Tr denotes the sum over all possible configurations of the dynamical variables
and $\beta=1/T$ is the inverse temperature. 
In the case of disordered system, 
one may evaluate $Z(\textbf{J})$ for quenched disorder $\textbf{J}$ and
take average of $\log Z(\textbf{J})$ over $\textbf{J}$ with an
appropriate weight. Using the replica method\cite{MPV1987}, 
the averaged free energy $[F]$ is rewritten as a limit of cumulant
generating function $\phi(n)$ of $F(\textbf{J})$ as 
\begin{equation}
 [F] = \lim_{n\to 0}\left\{-\frac{1}{N\beta n}\log[Z^n]\right\}
  =:\lim_{n\to 0} \phi(n),  \label{repid}
\end{equation}
where 
$[\cdots]$ denotes the average
with respect to the quenched disorder. 

 In case where $n$ is a real number, to proceed the calculation of the
 right hand side in Eq.~(\ref{repid}) needs some ansatz. A typical one is 
replica symmetric (RS) ansatz, which is considered to be correct only for
sufficiently large $n$. We denote the solution based on the RS ansatz as
 RS solution $\phi_{\rm RS}(n)$. 
Thus, the limit of $\phi(n)$ we are interested in
becomes nontrivial when we have no alternatives except the RS solution.

\begin{figure}[]
\includegraphics[width=\figwidth]{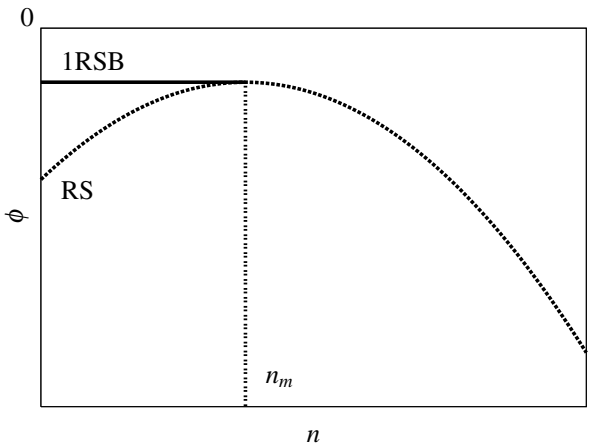}
\caption{
A schematic figure of $\phi(n)$ as a function of the replica number $n$.
This shows the construction of a 1RSB solution using monotonicity and
 convexity condition.
The dashed line represents RS solution, which breaks the monotonicity
 condition at $n=n_m$. Below $n_m$, $\phi(n)$ becomes a constant function
 down to zero, corresponding to the 1RSB solution. 
} 
\label{sch1rsb}
\end{figure}

In general, however, the function $\phi(n)$ is restricted by the following conditions:
$\phi'(n) \leq 0$(monotonicity), $(n\phi(n))'' \leq 0$(convexity), and 
AT stability. The two former conditions, monotonicity and convexity, come from a thermodynamic
restriction if the replica number $n$ is regarded as a ``temperature''. In
particular, they lead to the following proposition\cite{NH2008}: 
\begin{center}
if $\phi'(n_m)=0$ for $n_m>0$, \\
then $\phi(n)=\phi(0)$ for $0\leq n\leq n_m$.
\end{center}
Therefore, if the RS solution is valid for $n\geq n_m$, the limit $n\to
0$ is performed by this 
proposition. 
Figure \ref{sch1rsb} shows how the function $\phi$ is 
connected to the origin. 
It is also shown\cite{NH2008} that the solution $\phi_{\rm RS}(n_m)$ 
corresponds to the 1RSB solution for a wide class of models with $q_0=0$,
not restricted to the fully connected models. 
This relationship has already been pointed out in a solvable model\cite{OK2004}.

The proposition provides us a simple construction of a 1RSB solution using
only the RS solution. 
We summarize our procedure for the 1RSB construction as follows: 

\begin{enumerate}
 \item Calculate the RS solution $\phi_{\rm RS}(n)$ as a function of the
       finite replica number $n$. 
 \item Find the value $n_m$ which satisfies $\phi'_{\rm RS}(n_m)=0$.
 \item Set
\begin{equation}
\phi(0)=\phi_{\rm RS}(n_m) .
\label{nackofrep}
\end{equation}
\end{enumerate}
While the right hand side of Eq.~(\ref{nackofrep}) 
 is analytically tractable but doubtful for $n\ll 1$ because of
the RS ansatz, the left hand side is equal to the free energy as stated
 in  Eq.~(\ref{repid}) but analytically intractable.

One may notice that this procedure is analogous to the original
saddle-point method, if one identifies the replica number with
the breaking parameter. We consider this correspondence as the 
reason why we have to
maximize with respect to the breaking parameter in literatures.
It should be noted that this procedure can
apply to any model in which the RS solution is explicitly obtained for
any real $n$. 
Our procedure does not require
overlap matrix or the introduction of breaking parameter.

\subsection{Model}

Hereafter we deal with a finite-connectivity Ising SG model. The Hamiltonian
with $K$-spin interactions on a regular random graph with connectivity $C$ is
defined as: 
\begin{equation}
 H=-\sum_{\mu \in {\cal G}} {\cal D}_{\mu}J_{\mu} \sigma_{\mu(1)} \sigma_{\mu(2)}\cdots \sigma_{\mu(K)},
\end{equation}
where
\begin{eqnarray}
 {\cal G} &=& \biggl\{\mu=\{\mu(1),\cdots,\mu(K)\}; \; \nonumber\\
&\;&\mu(i)\in\{1,2,\cdots,N\},\mu(i)\neq\mu(j)
(i\neq j)\biggr\}.
\end{eqnarray}
Here $\sigma_i=\pm 1$ represents Ising spins on the random graph with
 $N$ sites. The interactions $J_{\mu}$ take $\pm 1$  with equal
 probability which gives the unit of energy and temperature. ${\cal
 D}_\mu=0,1$ are quenched variables, satisfying 
 the condition 
$\sum_{\mu\in{\cal G},i\in\mu}{\cal D}_\mu =C$ for each site $i$, namely
all the sites having the same number of the neighbors $C$.

\subsection{Solutions for 1RSB}

We calculate the cumulant generating function of the model described above within the framework of
Sec.~\ref{sec21}.  Following the calculation\cite{WS1987,Mim,Mim2}, $\phi(n)$
under the RS ansatz is evaluated as

\begin{figure}
 \includegraphics[width=\figwidth]{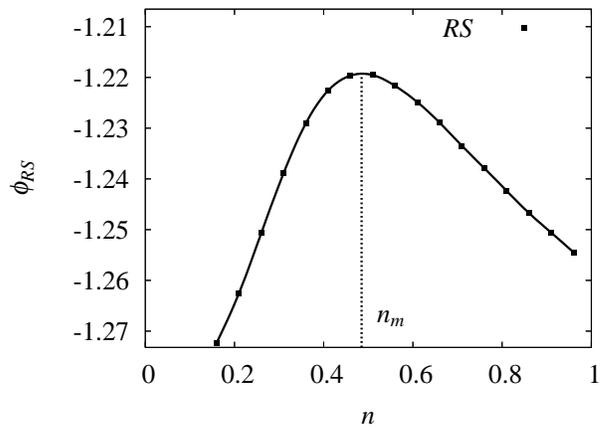}
\caption{Replica number $n$ dependence of $\phi_{\rm RS}(n)$ of a
 finite-connectivity Ising SG for $K=3$ and $C=4$ at $T=0.33$.}
\label{funcphi} 
\end{figure}

\begin{widetext}
\begin{equation}
 \phi_{\rm RS}(n)=-\frac{C}{K\beta}\log(\cosh(\beta))-
\frac{1}{\beta n}{\rm extr}_{\pi,\hat\pi}\left\{\frac{C}{K}\log
					     I_1-C\log I_2+\log
					     I_3\right\},
\label{eq:phiRS}
\end{equation} 
where
\begin{eqnarray}
 \displaystyle I_1&=&\int\prod_{k=1}^K {\rm
  d}x_k\pi(x_k)\;\frac{1}{2}\sum_{J_\mu=\pm 1}\left\{1+\tanh(\beta
  J_{\mu})\prod_{k=1}^K \tanh(\beta x_k) \right\}^n,\\
 \displaystyle  I_2&=&\int {\rm d}x{\rm d}\hat x \pi(x)\hat \pi(\hat x)\left\{1+\tanh(\beta
						x)\tanh(\beta  \hat x)\right\}^n,\\
 \displaystyle  I_3&=&\int\prod_{\gamma=1}^C {\rm d}\hat
  x_\gamma\hat \pi(\hat x_\gamma)\left\{\prod_{\gamma=1}^C(1+
				  \tanh(\beta \hat
				  x_\gamma))+\prod_{\gamma=1}^C(
				  1-\tanh(\beta \hat x_\gamma))
				 \right\}^n.
\end{eqnarray}
Differentiating $\phi_{\rm RS}$ with respect to $\pi$ and $\hat \pi$, we
 have the saddle-point equations 
\begin{eqnarray}
 \pi(x)&=&\frac{I_2}{I_3}\int\prod_{\gamma=1}^{C-1}{\rm d}\hat
  x_\gamma\hat \pi(\hat x_\gamma)\left\{\prod_{\gamma=1}^{C-1}(1+
  \tanh(\beta \hat x_\gamma))+\prod_{\gamma=1}^{C-1}
( 1-\tanh(\beta \hat x_\gamma))
				 \right\}^n\delta\left(x-\sum_{\gamma=1}^{C-1}\hat
				 x_{\gamma}\right),
\label{iter1}\\
\hat \pi(\hat x)&=&\frac{I_2}{I_1}\int \prod_{k=1}^{K-1} {\rm d}x_k\pi(x_k)
\;\frac{1}{2}\sum_{J_\mu=\pm 1}\delta\left(\hat x -\frac{1}{\beta}{\rm atanh}\left(\tanh((\beta
					    J_\mu)\prod_{k=1}^{K-1}\tanh(\beta x_k))\right)\right). \label{iter2}
\end{eqnarray}
\end{widetext}
We solve Eqs. (\ref{iter1}) and (\ref{iter2}) for each $n$ numerically
and obtain the saddle-point functions $\pi(x)$ and $\hat{\pi}(\hat x)$. 
Details for the numerical method we use to solve these equations are
shown in Appendix \ref{apppi}. 
Inserting the saddle-point functions into Eq.~(\ref{eq:phiRS}), 
we evaluate $\phi_{\rm RS}(n)$ as a function of $n$.
Fig.~\ref{funcphi} shows an example of $\phi_{\rm RS}(n)$ plotted against $n$ for
$K=3$ and $C=4$ at $T=0.33$, which is well below the expected SG
transition temperature, $T_c\approx 0.65$. 
As shown in the figure, $\phi_{\rm RS}(n)$ violates the monotonicity
condition at a certain value $n_m(T)$ which is defined by 
$\phi_{\rm RS}'(n_m)=0$. 

Following our scheme mentioned above, this is enough to
construct a 1RSB solution. 
The 1RSB free energy per site $f$ is given as $f=\phi_{\rm
RS}(n_m)$. 
It would be interesting to see the information of finite replica number is 
used to describe the 1RSB free energy. This is a consequence of 
the thermodynamic construction, with which the RS solution is connected to 
the physical limit $n\to 0$.

\begin{figure}
 \includegraphics[width=\figwidth]{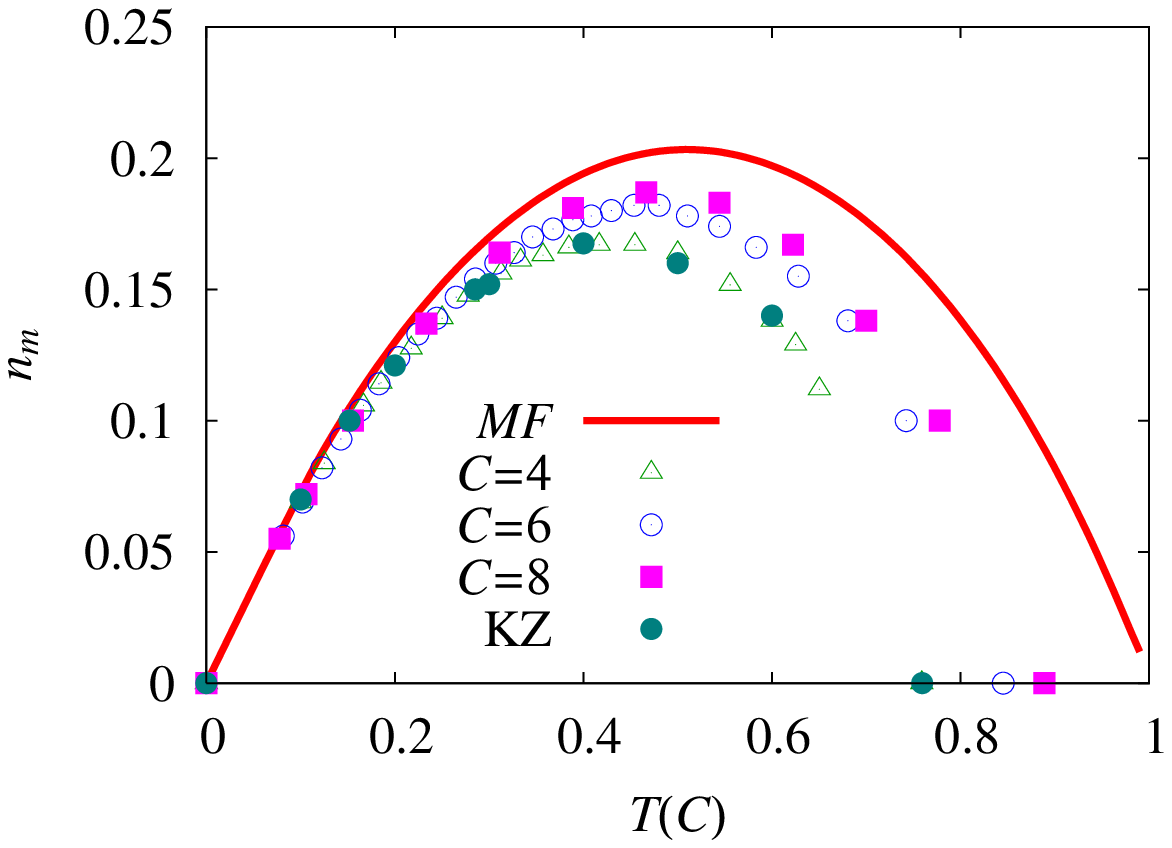}
 \includegraphics[width=\figwidth]{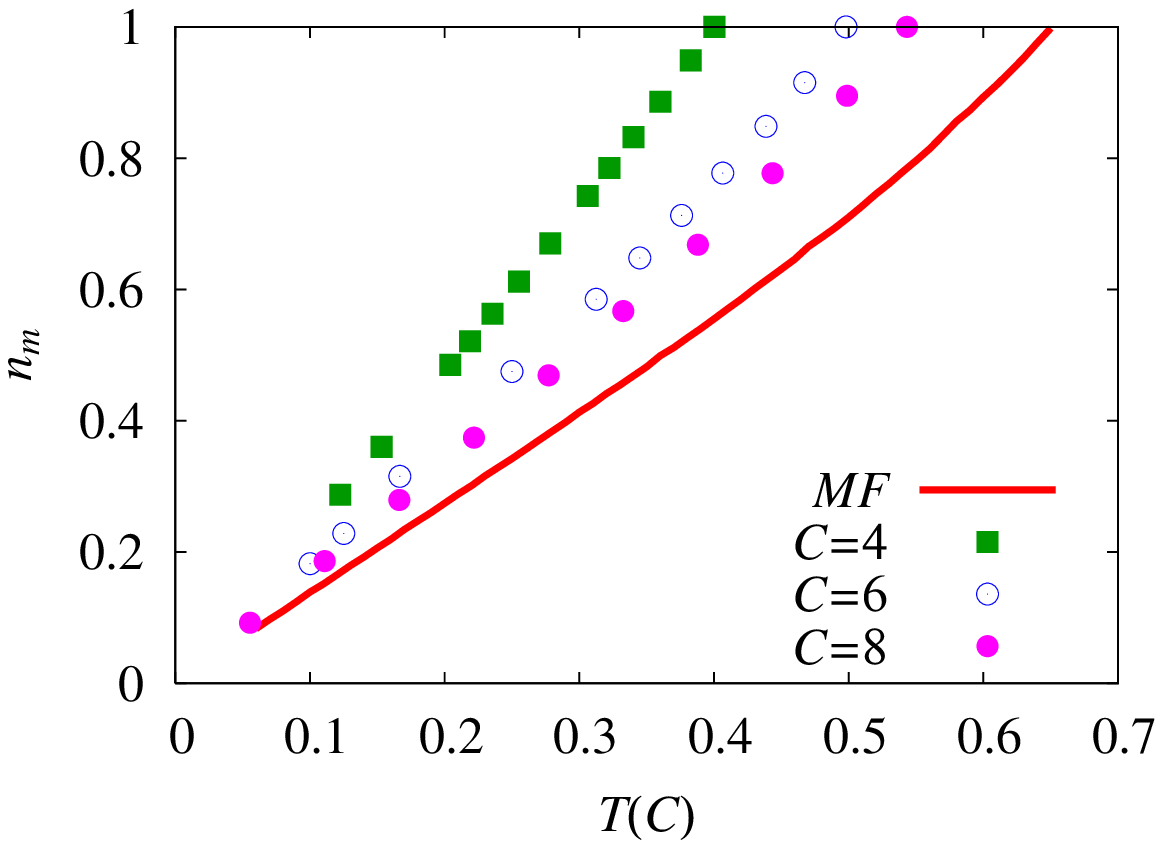}
\caption{
(Color online)
Temperature dependence of $n_m$ for $K=2$ (top panel) and
  for $K=3$ (bottom panel). Temperature is scaled as 
$T(C)=T\sqrt{K/2C}$. 
Each mark represents $n_m$ for connectivity $C=4$, $6$ and $8$. 
The solid line represents $n_m$ for $K$-body  Sherrington-Kirkpatrick
 model with Gaussian interaction.   
In the top panel, thermodynamic value
of 1RSB parameter for $K=2$ and $C=4$ evaluated in Ref. \onlinecite{KZ2008} 
is also shown in filled circle.
}
\label{fig-nm}
\end{figure}

We have evaluated $\phi_{\rm RS}(n)$ at $0\leq n\leq 1$ 
for $K=2$ and $K=3$,
which yields temperature dependence of the 1RSB free energy shown later. 
For comparison, we also evaluate an RS free energy, which is defined as
$\phi_{\rm RS}(0)$. 
Temperature dependence of $n_m$ for some values of $C$ is plotted for
$K=2$ and $3$ in Fig. \ref{fig-nm}. 
We also show the parameter $m$ for $K=2$ and
 $C=4$ in Fig.~\ref{fig-nm}, evaluated in Ref.~\onlinecite{KZ2008}. They are
 in good agreement with each other.
 The transition temperature for $K=2$ is derived from the condition that
 the instability condition of $\pi(x)=\delta(x)$ and then $n_m$ begin to
 deviate from zero.  The estimate of $T_c$ is consistent with the known
 expression $T_c= 1/{\rm atanh} (C-1)$\cite{T1986} 
 considering an appropriate factor $\sqrt{1/C}$. For $K=3$, $T_c$ is
 determined by an onset temperature at which the monotonicity breaking
 point emerges. Then, $n_m$ deviates from unity, that is often observed
 in some models exhibiting 1RSB transition. While the analytic
 expression of $T_c$ for $K=3$ has not known yet, the estimate for
 $C=4$ and $8$ is consistent with that obtained by the cavity method\cite{MR2004}.

Here we compare our scheme to the established cavity method, in
particular for the finite connectivity Ising SG model\cite{MR2004}. 
The saddle-point equations, Eqs. (\ref{iter1}) and (\ref{iter2}), in our
scheme are the same as the recursion equation derived as Eqs. (A.3) and
(A.4) in Ref. \onlinecite{MR2004}, when the functions $\pi$ and
$\hat{\pi}$ are identified as the distribution of cavity field and
cavity bias, respectively. While the parameter $n$ is determined by the
monotonicity condition $\phi'(n)=0$, the 1RSB parameter $m$ in the
cavity context is determined by the non-negativity condition of the
complexity $\Sigma$:  
\begin{equation}
 \Sigma(f(m))=\beta m^2 \phi'(m) =0
\end{equation}
within the formalism of Monasson\cite{M1995,P2003}. This means that these two methods are
equivalent when the complexity is a well-defined quantity.

In the previous works\cite{MP2001,FLRZ2001,MR2004}, 
it is shown that the result of the cavity method corresponds to that of
the replica method with a factorized ansatz for the
finite connectivity models. Thus, our construction 
is also equivalent to the replica theory with the  factorized ansatz.
In the formalism,
the replica number $n$ is substituted for the breaking parameter $m$
in the expression of free energy without taking the limit $n\to 0$.
Then, the maximization of the free energy with respect to the overlap
parameter $q$ and breaking parameter $m$ is equivalent to 
the monotonicity breaking condition in our scheme. 
This reasoning does not give a correctness proof of the factorized ansatz
(and also our) solution,  
but we convince ourselves that it reveals the reason why the factorized
ansatz gives numerically correct solution.

\section{Verification by Numerical Simulation}
\label{sec:num}
\subsection{Monte Carlo method}

In the previous section, we obtain the 1RSB solution for 
the Ising SG model  with $K$-body interactions by using our scheme. This
is the true solution if 
the AT instability or others would not occur above $n_m$, but it is
difficult to examine the validity of $\phi_{\rm RS}(n)$. 
This situation is similar to the case of the cavity method. 
Instead, here we verify our 1RSB solution by comparing it to Monte Carlo (MC)
data. We use exchange MC method\cite{HN1996} in order to accelerate
relaxation time to equilibrium.  
The number of temperatures is fixed to be 30 and the lowest
temperature is down to 0.5 for all the system sizes $N$ and $K$.
The simulation parameters for $K=2$ and $K=3$ are presented in Table 
\ref{24mcs} and \ref{34mcs}, respectively. 
Equilibration of the MC simulations is confirmed by seeing that the
observed quantities are stable within range of error by doubling MC
steps. 

By using the MC simulation 
we measure the energy $e_N(T)$ per site and calculate the free energy $f_N(T)$
per site by  thermodynamic integration: 
\begin{equation}
 f_N(T)=T\int_T^\infty {\rm d} T' \frac{e_N(T')}{T'^2},
\end{equation}
and the entropy $s_N(T)$ per site as 
\begin{equation}
 s_N(T)= \frac{e_N(T)-f_N(T)}{T}.
\end{equation}
Through the data at discrete temperatures obtained by the exchange MC method, 
the energy as a continuous function of $T$ is evaluated by 
reweighting formula\cite{FS1989}:
\begin{equation}
 \langle A({\bf \sigma})\rangle_{MC}^{(\beta)} 
= \frac{\left\langle A({\bf \sigma}) 
{\rm e}^{(\beta_0-\beta)H({\bf \sigma})}\right\rangle_{MC}^{(\beta_0)}}
{\left\langle{\rm e}^{(\beta_0-\beta)H({\bf \sigma})}
\right\rangle_{MC}^{(\beta_0) }}, 
\end{equation} 
where $\langle\cdots\rangle_{MC}^{(\beta)}$ denotes the MC average at
the inverse temperature $\beta$. We apply this formula by setting  
$\beta_0$ as actually simulated temperature and $\beta$ as required one. 
We choose $\beta_0$ as the nearest temperature to 
 $\beta$ from the whole set of simulated temperatures.

\begin{table}
\begin{tabular}{|c|c|c|}
\hline
$N$& $N_{\rm MCS}$ & $N_{\rm s}$\\
\hline
32 & $1\times 10^5$ & 4096\\
48 &$2\times 10^5$& 2048\\
64&$4\times 10^5$& 1024\\
128&$3\times 10^6$& 512\\
256&$5\times 10^7$& 128\\
512&$5\times 10^8$& 30\\
\hline
\end{tabular}
\caption{
Parameters of simulation in the case of $K=2$ and $C=4$.  
The total number of Monte Carlo steps $2N_{\rm MCS}$ 
and the total number of samples $N_s$  are presented for each size $N$. 
The first $N_{\rm MCS}$ are discarded for equilibration and the subsequent
$N_{\rm MCS}$ are used in measurement. 
} 
\label{24mcs}
\end{table}
\begin{table}
\begin{tabular}{|c|c|c|}
\hline
$N$ & $N_{\rm MCS}$ & $N_{\rm s}$\\
\hline
30 &$1\times 10^5$ & 4096\\
36 &$1\times 10^5$ & 4096\\
45 &$2\times 10^5$& 2048\\
60&$4\times 10^5$& 1024\\
75&$8\times 10^5$&1024\\
120&$3\times 10^6$& 512\\
240& $1\times 10^8$& 128\\
\hline
\end{tabular} 
\caption{
Parameters of simulation in the case of $K=3$ and $C=4$. 
The total number of Monte Carlo steps $2N_{\rm MCS}$ 
and the total number of samples $N_s$ are presented for each size $N$. 
The first $N_{\rm MCS}$ are discarded for equilibration and the subsequent
$N_{\rm MCS}$ are used in measurement. 
}  
\label{34mcs}
\end{table}

\subsection{Results}

\begin{figure}[t]
  \includegraphics[width=\figwidtha]{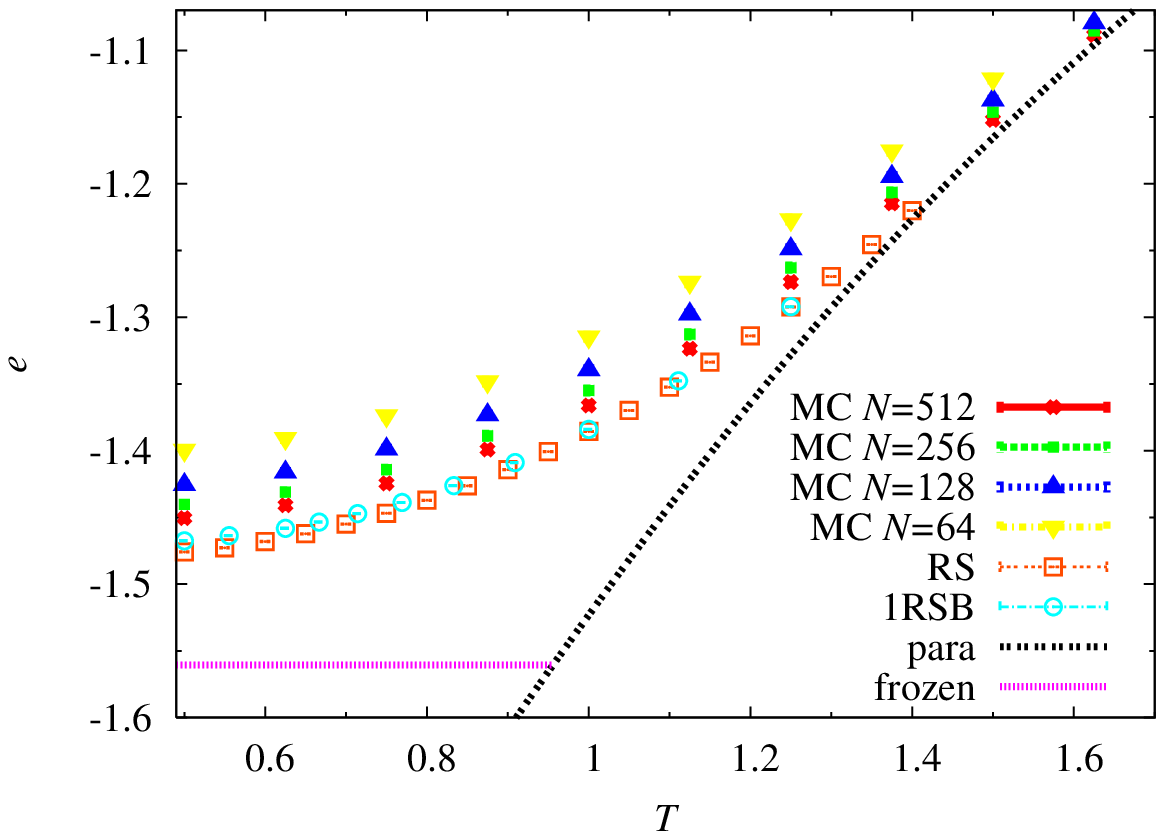}
  \includegraphics[width=\figwidtha]{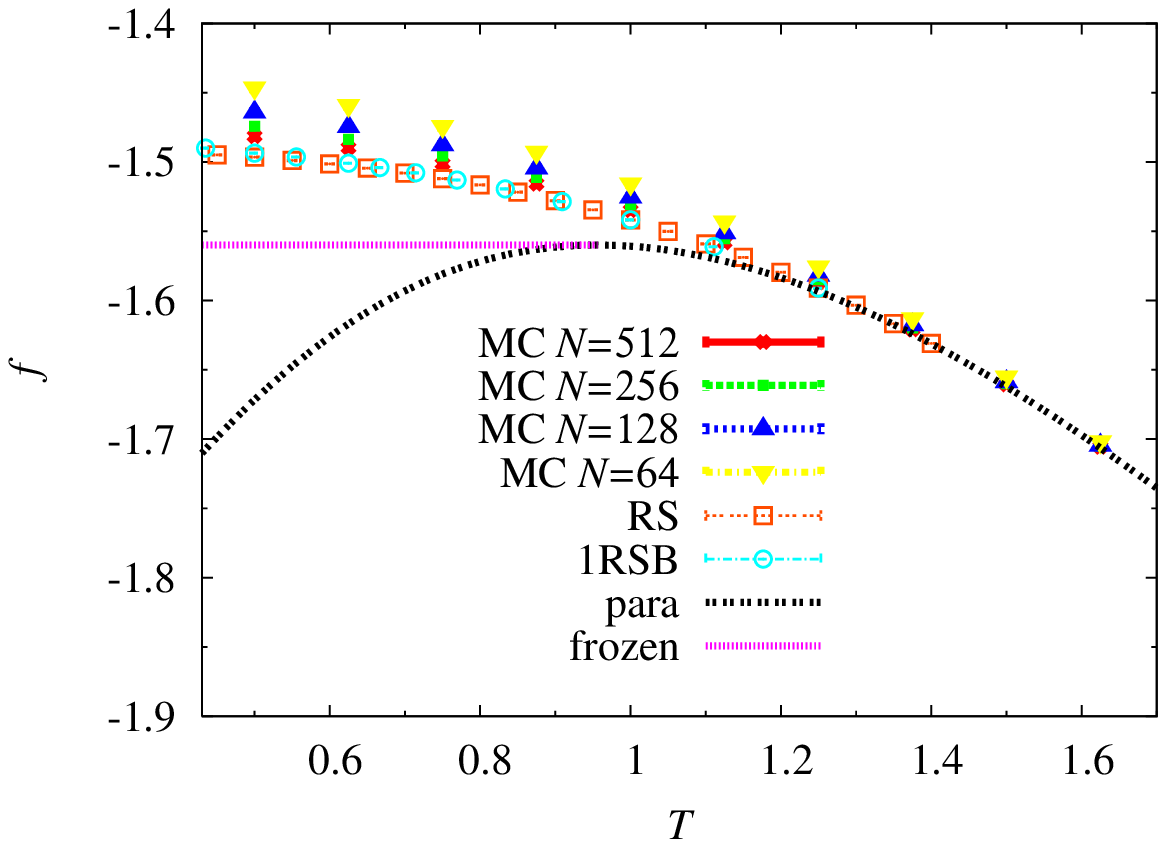}
  \includegraphics[width=\figwidtha]{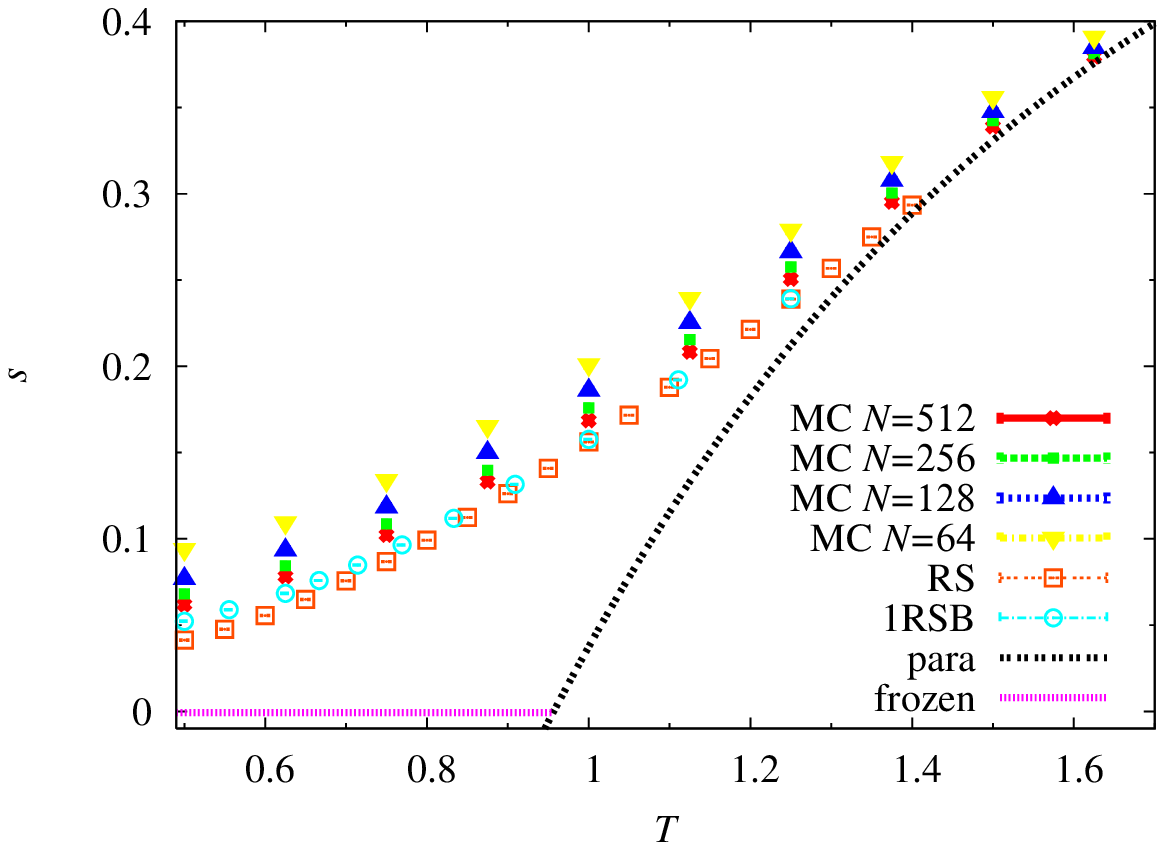}
  \caption{(Color online) Temperature dependence of energy(top panel),
 free energy(middle panel) and entropy(bottom panel) for a finite
 connectivity Ising SG with $K=2$ and  $C=4$.
MC results are shown by filled marks for $N=64$, $128$, $256$ and $512$
 from the top. Open squares and open circles are the results of the 1RSB
 solution and the RS one, respectively. 
The paramagnetic solution is presented by the dotted line and the frozen
 ansatz is solid line. 
}
\label{en24}
\end{figure}

\begin{figure}[t]
  \includegraphics[width=\figwidtha]{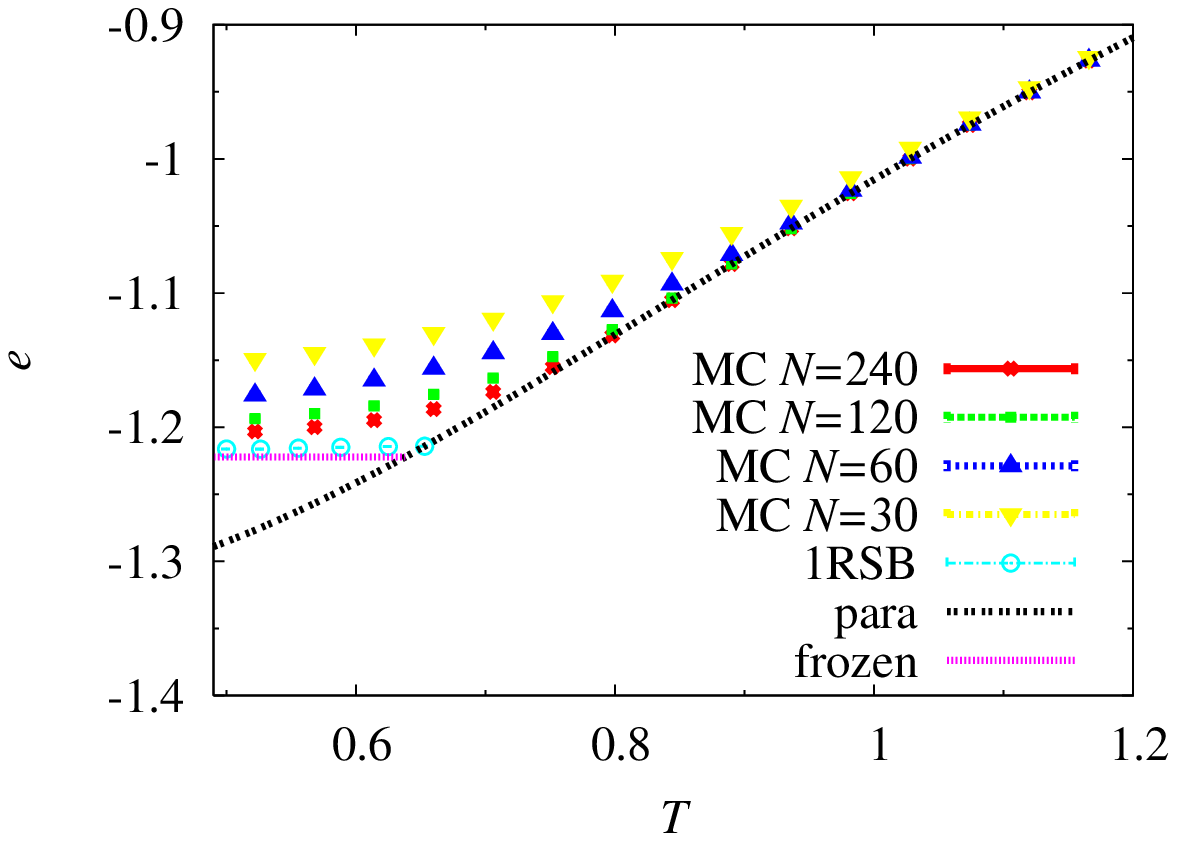}
  \includegraphics[width=\figwidtha]{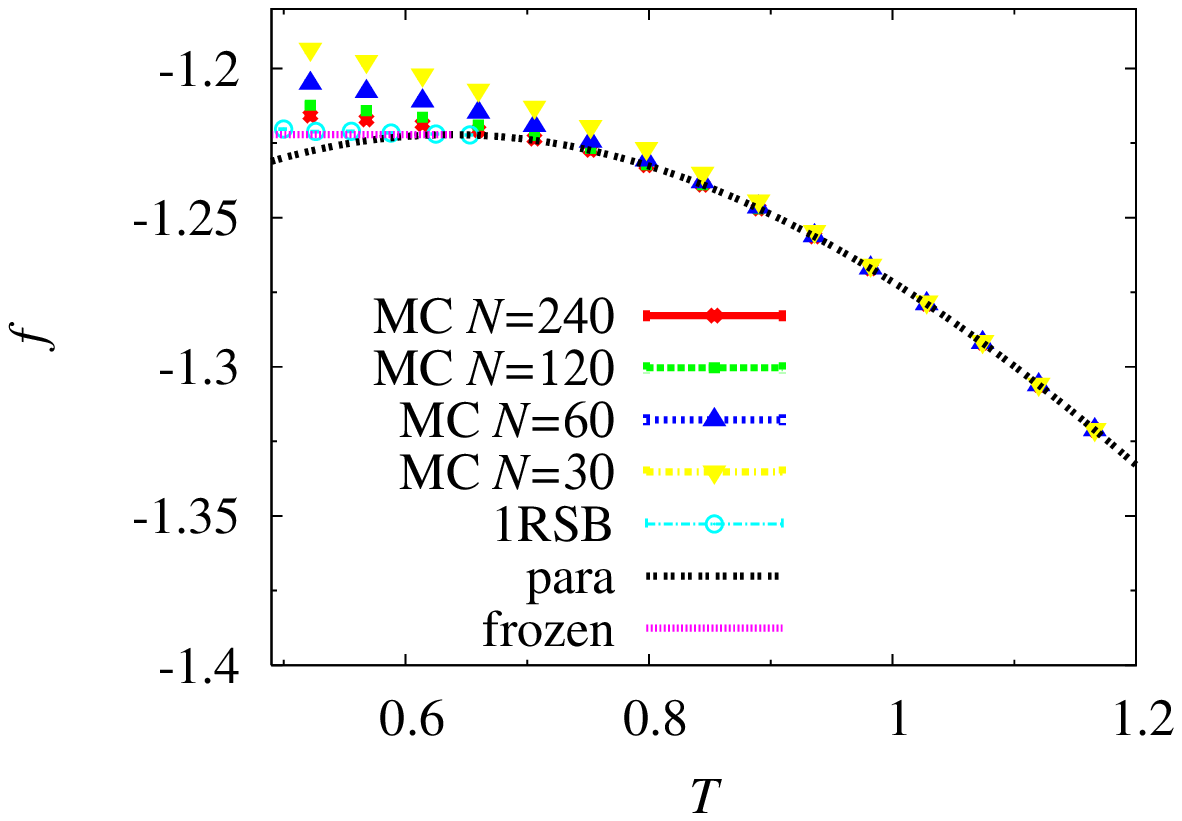}
  \includegraphics[width=\figwidtha]{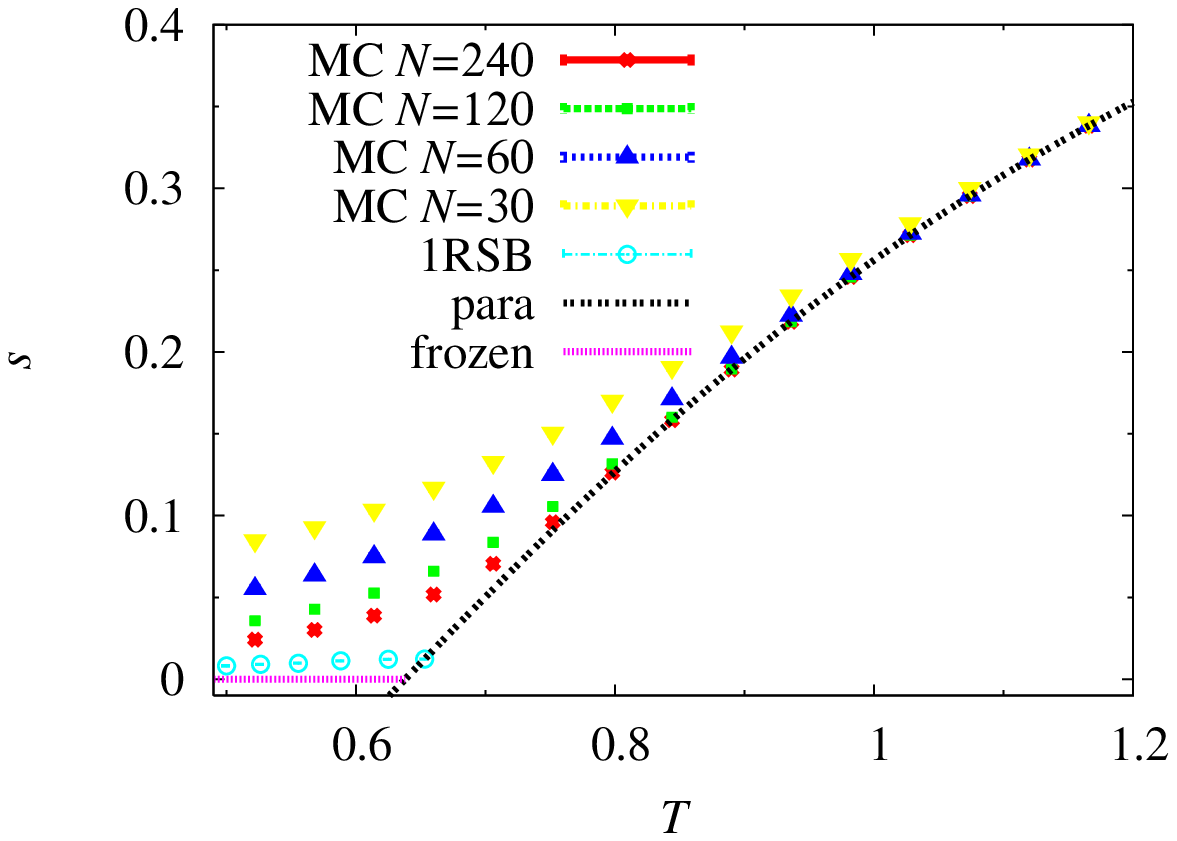}
  \caption{(Color online)
Temperature dependence of energy(top panel), free energy(middle panel)
 and entropy(bottom panel) for a finite-connectivity Ising SG with
 $K=3$ and $C=4$. 
MC results are shown by filled marks for $N=30$, $60$, $120$ and $240$
 from the top.
 Details of the lines are the same as those of Fig. \ref{en24}.}
\label{en34}
\end{figure}

\begin{figure}
  \includegraphics[width=\figwidtha]{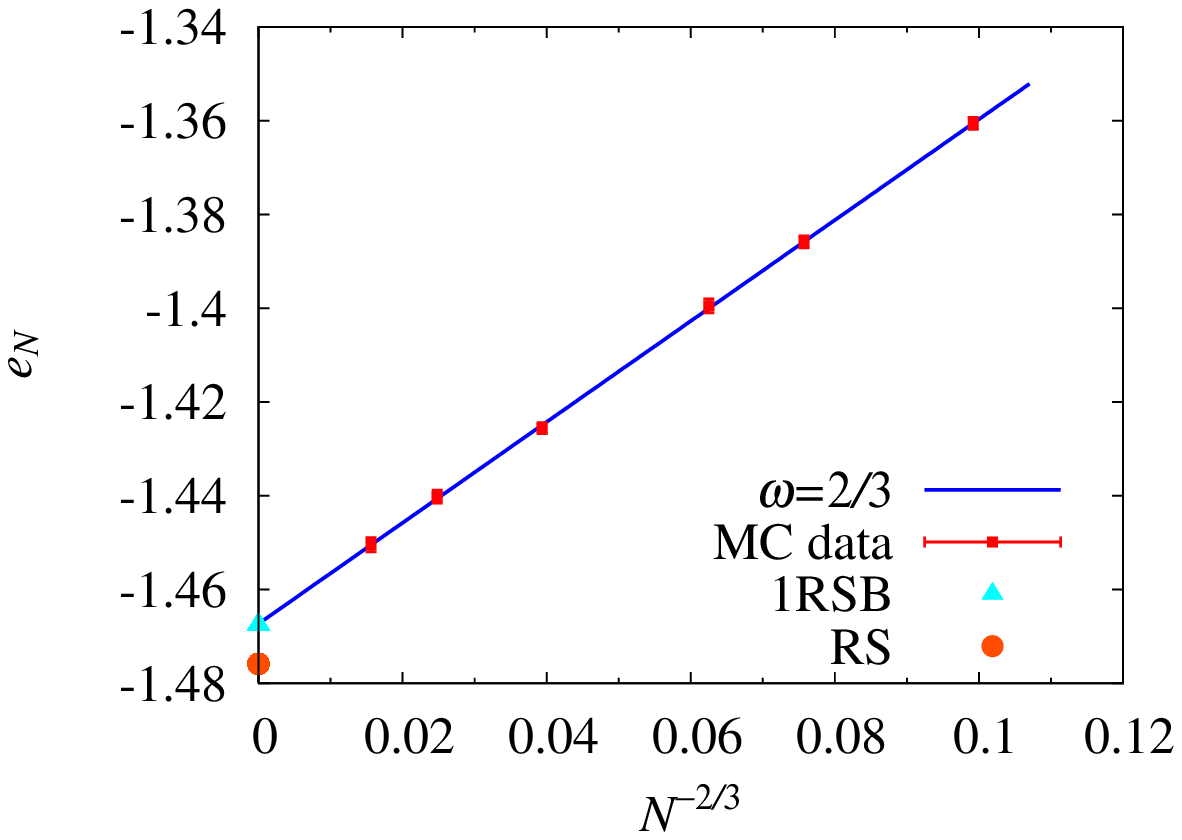}
  \includegraphics[width=\figwidtha]{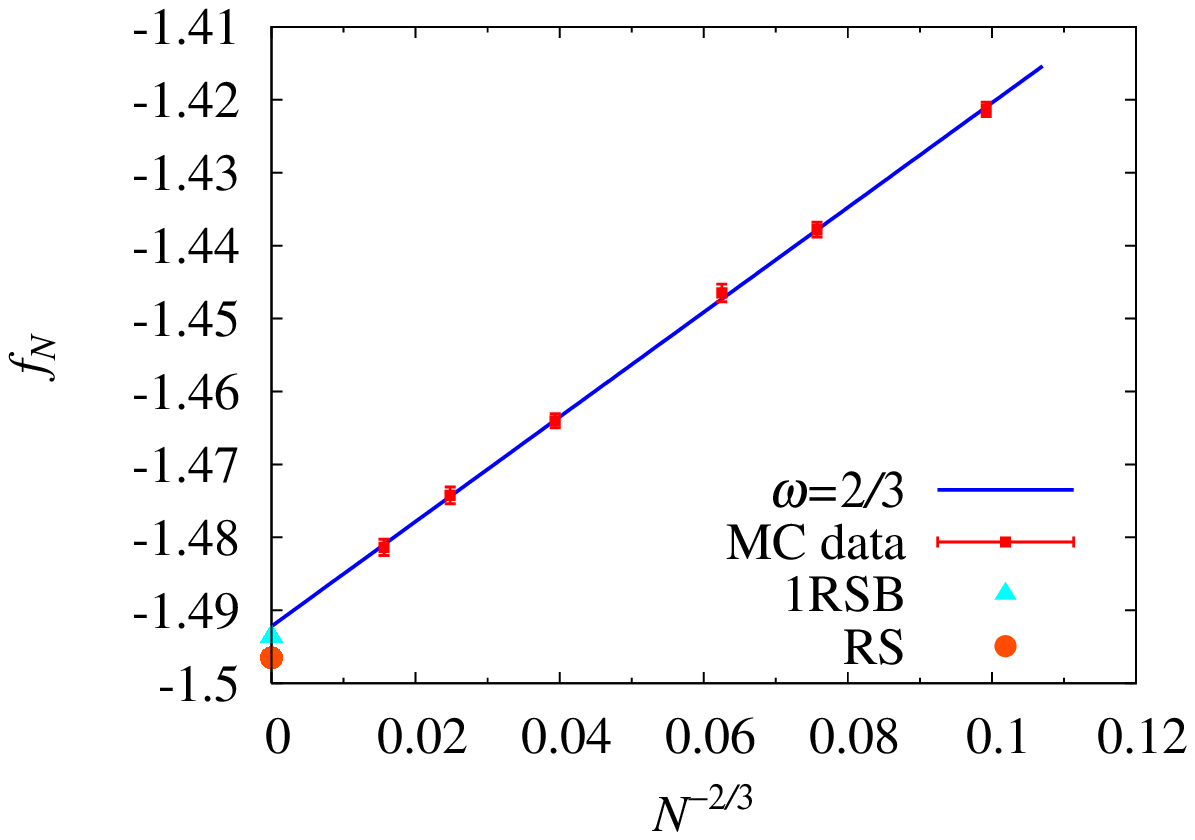}
  \includegraphics[width=\figwidtha]{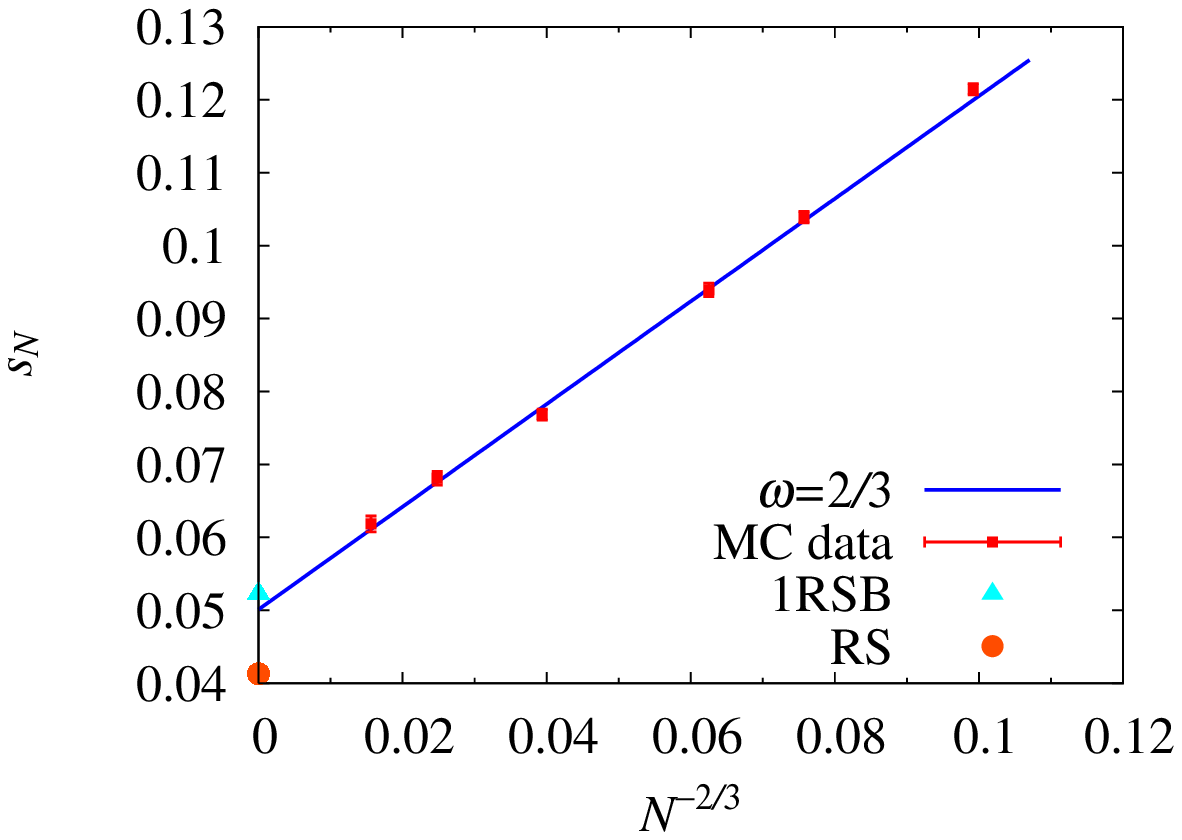}
  \caption{(Color online) Energy, free energy and entropy as a function
 of $1/N^{2/3}$ for a finite-connectivity Ising SG with $K=2$ and $C=4$ at $T=0.5$. 
The filled squares are MC results, filled triangle is 1RSB solution, and
filled circle is RS solution. In solid line,  
the least square fitting of MC results
 assuming the exponent of the leading finite-size correction $\omega$ is
 2/3 is presented.
}
\label{enext24}
\end{figure}

\begin{figure}[t]
  \includegraphics[width=\figwidtha]{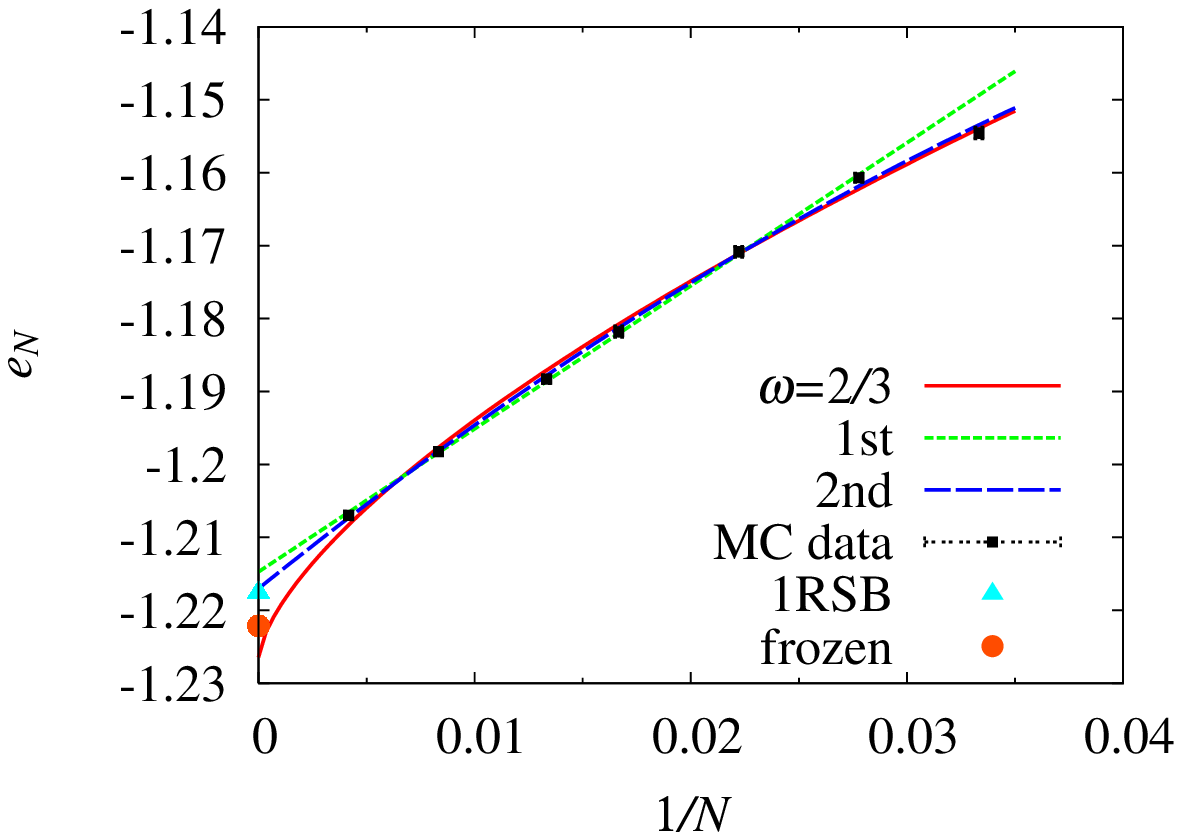}
  \includegraphics[width=\figwidtha]{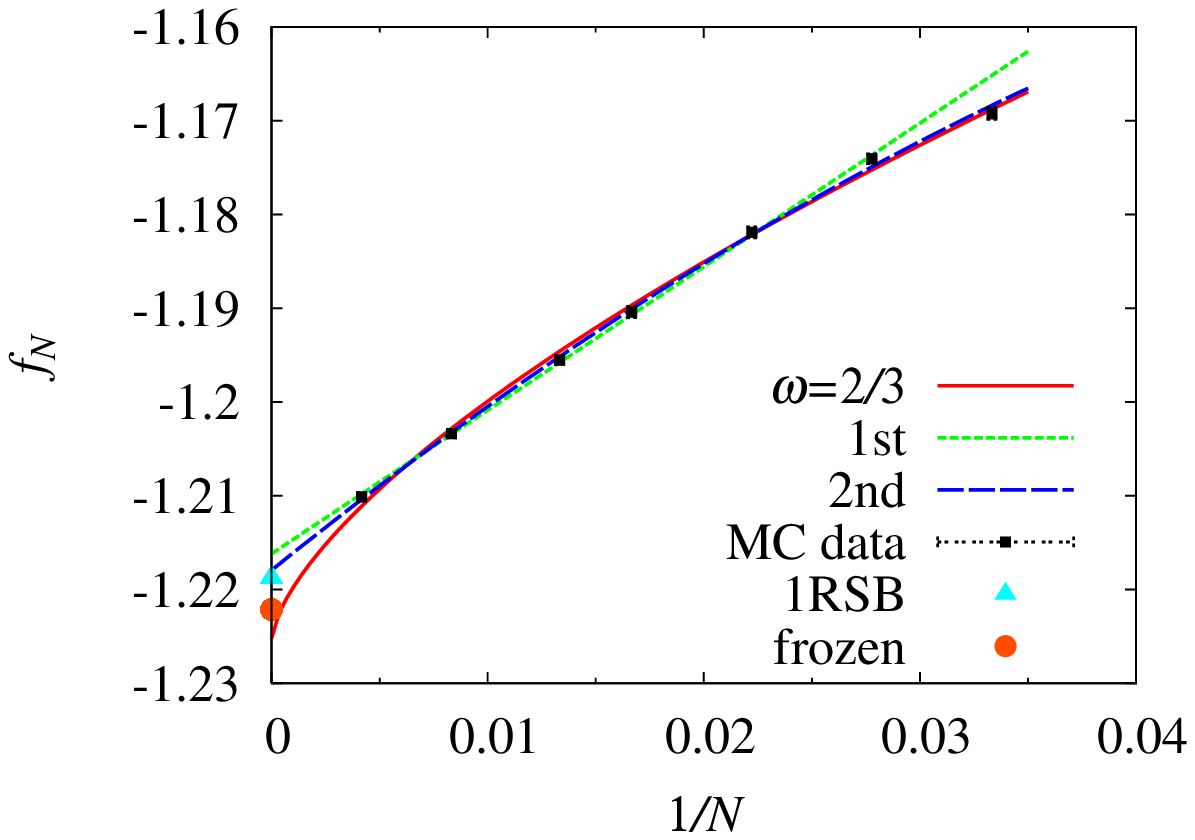}
  \includegraphics[width=\figwidtha]{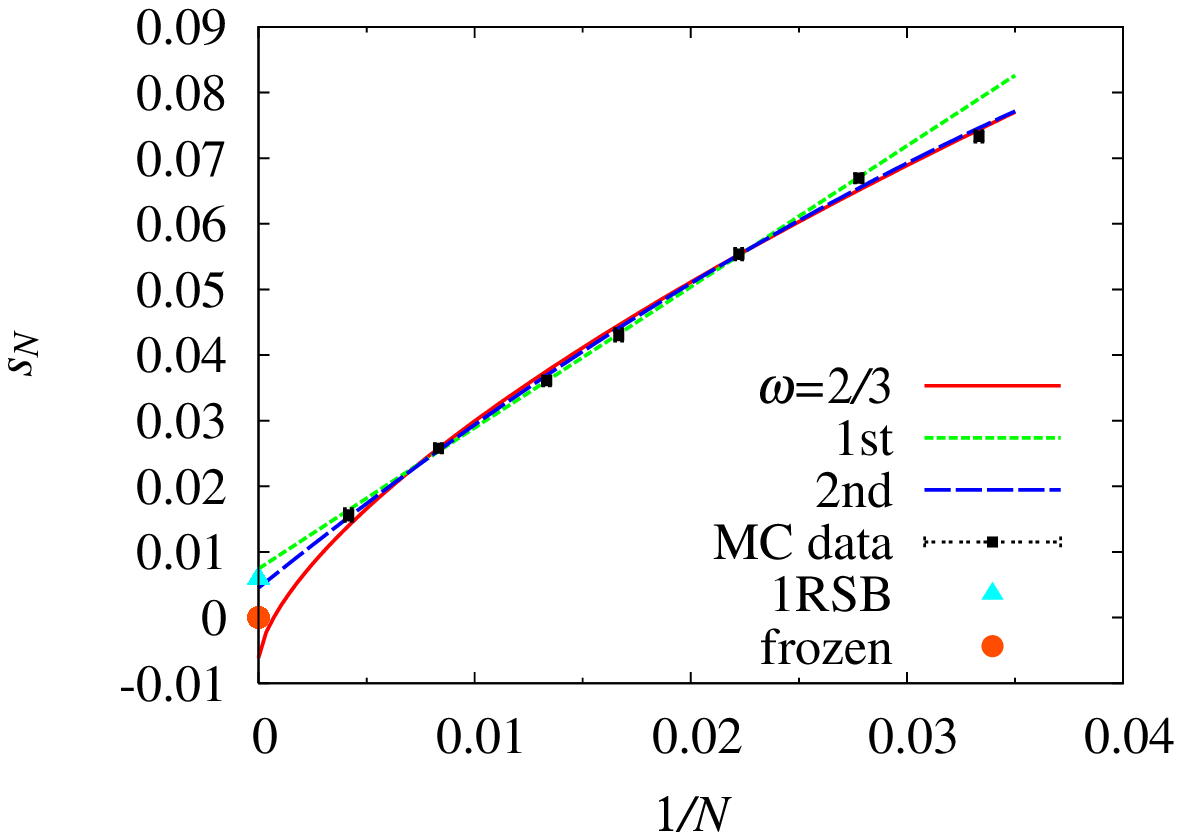}
  \caption{(Color online) Energy, free energy and entropy as a function
 of $1/N$ for 
a finite-connectivity Ising SG with 
$K=3$ and $C=4$ at $T=0.2$. 
The points and solid line are the same as in Figs.~\ref{enext24}.
Short-dashed and long-dashed lines represent the least-squares fits for
 the form including the leading term of $1/N$ and up to the next leading
 terms, respectively. 
}
\label{enext34}
\end{figure}

We display thermodynamic quantities, energy, free energy and entropy,
obtained by MC simulations, together with the RS and 1RSB solutions 
for $K=2$ and $C=4$ in Fig.~\ref{en24} and for $K=3$ and $C=4$ 
in Fig.~\ref{en34}. 
The data show that in the case of $K=2$, the RS and 1RSB solutions are close to
each other, but
the 1RSB free energy is always greater than the RS one by definition.

For $T>T_c$, the correct solution is given by the paramagnetic one,
which is described by $\pi(x)=\hat{\pi}(x)=\delta(x)$. For comparison, we
also plot a practical solution based on frozen ansatz\cite{MO2003}, 
in which the paramagnetic solution is used at $T>T_g$
and the entropy is kept to zero at $T<T_g$. Here $T_g$ is defined as
the temperature at which the entropy given by the paramagnetic solution
is zero.  This ansatz leads to the results that the free energy as well as the energy
is independent of $T$ below $T_g$. 
The frozen ansatz is interpreted as a paramagnetic solution on which the
monotonicity condition as a function of temperature is imposed. Although
the true free energy must be a monotonically decreasing function of
temperature in a standard thermodynamic sense, it might not be
a sufficient condition.  
In fact, MC data and the 1RSB solution are far from the frozen-ansatz
solution. In particular, they show non-zero value of the entropy at
finite temperatures as shown in the bottom of Fig.~\ref{en24}, which is
quite different from that of the frozen ansatz.

For $K=3$, we do not plot the RS solution because we cannot find it
at low temperatures near $n=0$ except for the paramagnetic one. 
This suggests that the 1RSB scheme works even if the RS solution does not
exist near $n=0$, though we cannot rule out the possibility that our algorithm for
evaluating $\pi$ is unstable to find the RS solution.

\begin{table}
  \begin{tabular}{|c|c|c|c|}
 \hline
  & MC & 1RSB & RS\\
 \hline 
 $e_\infty$ & -1.4673(4)&-1.4675(2)&-1.4758(2)\\
 $f_\infty$& -1.4922(4) & -1.4937(1)&-1.4965(1)\\
 $s_\infty$  &0.0501(9) & 0.0522(2)&0.04128(2)\\
 \hline
  \end{tabular}
\caption{
Thermodynamic limit of the energy, free energy and entropy of a
 finite-connectivity Ising SG for $K=2$ and $C=4$ at $T=0.5$.
1RSB and RS represent those estimated from the 1RSB 
solution and the RS one, respectively. MC means 
the extrapolated values from finite-size MC
 data by assuming a power law of the leading correction with the
 exponent $\omega=2/3$.    
}
\label{bestfit}
\end{table}

\begin{table}
  \begin{tabular}{|c|c|c|c|}
 \hline
  & MC & 1RSB & frozen ansatz\\
 \hline 
 $e_\infty$ & -1.217(1) &-1.2176(1)&-1.2221\\
 $f_\infty$& -1.2180(8) & -1.2188(1)&-1.2221\\
 $s_\infty$  &0.005(2) & 0.0058(5)&0\\
 \hline
  \end{tabular}
\caption{
Thermodynamic limit of the energy, free energy and entropy of a
 finite-connectivity Ising SG for $K=3$ and $C=4$ at $T=0.2$.
1RSB and frozen ansatz represent those estimated from the 1RSB 
solution and the frozen-ansatz one. MC means 
the extrapolated values from finite-size MC
 data by assuming a form of  $x_N=x_{\infty}+a_1 N^{-1}+ a_2N^{-2}$, 
where $x=e,f \mbox{ or } s$.
}
\label{bestfitk3}
\end{table}

To see thermodynamic properties, we extrapolate our MC data with finite
sizes to the thermodynamic limit 
$N\to \infty$. 
Because finite-size correction terms 
and its exponent are a priori
unknown in SG models, an extrapolation method 
itself should be investigated.
We assume that the leading finite-size correction terms for the energy, free
energy and entropy are expressed as 
\begin{eqnarray}
  e_N=e_\infty+a_e\times N^{-\omega},\label{eqn:EFenegy}\\
  f_N=f_\infty+a_f\times N^{-\omega},\label{eqn:FFenegy}\\
  s_N=s_\infty+a_s\times N^{-\omega}, \label{eqn:SFenegy}
\end{eqnarray}
where $e_\infty$, $f_\infty$ and $s_\infty$ are the thermodynamic limit
of the respective quantities, and 
the correction exponent $\omega$ is assumed to be independent of the
quantities. 

\begin{figure}
 \includegraphics[width=\figwidth]{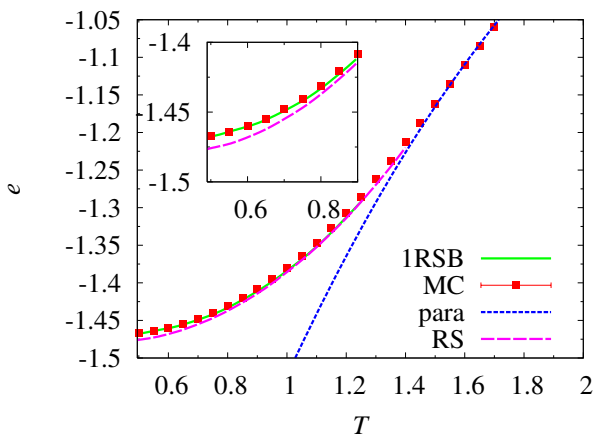}
\caption{(Color online)
Temperature dependence of the energy for a finite-connectivity Ising SG
 for  $K=2$ and $C=4$. The extrapolated value from MC data is
 marked by filled square. The 1RSB, RS and paramagnetic solutions are
represented  by solid, long-dashed and short-dashed lines,  respectively. 
The inset is an enlarged view at low temperatures. 
}
\label{k2temp} 
\end{figure}

As shown in the previous work\cite{B2003}, the ground-state energy of the
Ising SG model for $K=2$ defined on a regular random graph is scaled
with $\omega=2/3$. Thus,  we assume that the exponent $2/3$ holds for
$K=2$ at finite temperatures and is independent of physical quantities. 
Figs.~\ref{enext24} show the thermodynamic quantities as  a function of
$N^{-2/3}$ for $K=2$ at $T=0.5$, which is the lowest observed temperature. 
The data are fitted well with the assumption $\omega=2/3$ as
shown in the figure. 
The extrapolated values by the best fit and the results by the 1RSB and the
RS solutions are shown in Table \ref{bestfit}. 
The thermodynamic values by MC results agree with those by the 1RSB
solution rather than the RS one. The energy extrapolated  
in a wide range of temperature is displayed in
Fig.~\ref{k2temp}. 
This also suggests that the 1RSB solution is consistent with numerical
results. 

\begin{figure}
  \includegraphics[width=\figwidth]{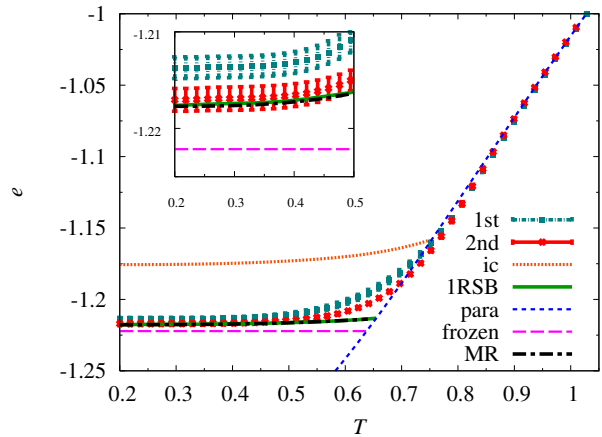}
\caption{(Color online) 
Temperature dependence of the energy for a finite-connectivity Ising SG
 for $K=3$ and $C=4$. 
The filled squares and cross marks are the extrapolated value from MC
 data with the extrapolation form including the leading correction term
 and up to the next correction terms, respectively. 
  The 1RSB, frozen and paramagnetic solutions are
represented  by solid, long-dashed and short-dashed lines,
 respectively. Long-short-dashed line is the result of the 
cavity method in Ref.~\onlinecite{MR2004}. 
The iso-complexity energy
obtained in Ref.~\onlinecite{MR2004} is also shown by dotted line.
The inset is an enlarged view at low temperatures. 
}
\label{k3temp}
\end{figure}

We turn to the case of $K=3$, where the value of $\omega$ is not known 
even at zero temperature.
Although a naive way to suppress 
higher order corrections is to study the
 system for large sizes and/or at lower temperatures apart from critical
 temperature, it has not been feasible to perform the MC simulation
 below $T=0.5\sim 0.7T_c$ for $N\ge 240$ in moderate CPU time 
 because of extremely slow relaxation especially in the case of $K=3$. 
This is contrast to $K=2$ model. 
However, for relatively smaller systems, the distribution
function of the energy is found to be almost 
a delta function with the weight at the lowest energy. This implies that 
vthe distribution depends weakly only on temperature $T$ below $0.5$.
This fact enables us to obtain the energy at temperatures down to $T=0.2
\sim 0.3 T_c$ using the reweighting method\cite{FS1989}. 
We evaluate the correction exponent $\omega$ for the energy by the
least-squares estimation at $T=0.2$ and $0.5$ with a form of
Eq.~(\ref{eqn:EFenegy}). 
The estimate of $\omega$ is not compatible with $\omega=2/3$ used in the
case of $K=2$, and is rather close to $\omega=1$. This tendency is
enhanced by omitting the smallest size $N=30$ from the analysis. These
findings suggest that $\omega\simeq 1$ and higher order corrections are
not negligible.

Therefore, we extrapolate the MC result for $K=3$ by assuming the forms
of Eqs.~(\ref{eqn:EFenegy}), (\ref{eqn:FFenegy}) and (\ref{eqn:SFenegy})
for $\omega=1$ with the next leading correction term $1/N^2$. 
The data for $N=30$ are omitted from the extrapolation analysis. 
Figures~\ref{enext34} shows the result of the thermodynamic quantities
for $K=3$ and $C=4$ at $T=0.2$. The extrapolated values, presented in
Table \ref{bestfitk3}, are consistent with those of the 1RSB solution by
taking into account the next leading correction term. 

We  also show the thermodynamic value of the energy for $K=3$ 
as a function of $T$ 
in Fig.~\ref{k3temp}.
The extrapolated values by the form including the next leading
correction term are 
consistent with those by the paramagnetic solution at $T>0.9$ and those
by the 1RSB solution at low temperatures, 
though a systematic deviation still remains around $T_c$ because of the
critical fluctuation. 
As shown in the
 inset of Fig.~\ref{k3temp}, the agreement between the extrapolated
 value and the value of 1RSB solution is held at very low temperatures and
the limiting value of energy at zero temperature coincides with that
 obtained by zero-temperature calculations\cite{FLRZ2001,MR2003,MP2004,LG1990-2}. 
For the case of $K=3$, the result of the cavity method is also
shown in Fig. \ref{k3temp}\cite{MR2004}. Analytic results are in good
agreement with the MC data at low temperatures. 
These support the validity of the scheme also for $K=3$.

Before closing this section, we would like to mention MC algorithm for
studying SG models. 
In recent works\cite{MR2004,KTZ2008}, it is claimed that 
in annealing simulations a slow annealing limit of the energy 
often leads to the iso-complexity energy, significantly above the static
equilibrium energy in glassy systems. 
This has been confirmed for $K=3$ by an annealing
simulation\cite{MR2004}. 
In contrast, as shown in Fig.~\ref{k3temp}, the energy extrapolated to
the infinite-volume limit in our exchange MC results is well lower than 
the iso-complexity energy and is compatible with that of the 1RSB
solution at low temperatures. 
This suggests that the exchange MC is suitable for equilibration of the
SG system even when the system have the iso-complexity energy separated
from the static one.

\section{Summary and Discussion}
\label{sec:sum}

We have studied a construction of a 1RSB solution for quench
disordered systems. Our construction is based on thermodynamic
conditions for the cumulant generating function $\phi(n)$ of free
energy,
which are derived as a necessary condition in the replica
analysis. 
The only requirement for our construction is to obtain the
replica symmetric solution for $\phi(n)$ as a function of $n$. 
This is a quite general scheme which may provide an unified way to give
a correct solution for 1RSB systems. In fact, our scheme reproduces the
well-known 1RSB solution for fully connected mean-field SG models such
as $p$-spin model\cite{NH2008} and Potts glass model\cite{NH2}. 
As a non-trivial
example we have applied our scheme to study a 1RSB solution for
finite-connectivity Ising SG models with $K$-spin interactions.
The thermodynamic quantities are explicitly evaluated from numerically
obtained RS solution with finite replica number $n$ using our
scheme. 

The saddle-point equations to be solved in our scheme are found to be
equivalent to recursion equations of the cavity-field distributions in
the 1RSB cavity formalism for this model. In a sense, our scheme
based on the replica theory can be regarded as a reinterpretation of the
1RSB cavity method.  
While the cavity method
can predict the microscopic detail of a model through
complexity,  
which is 
an interesting quantity in glassy physics, 
one cannot obtain such a quantity with our scheme at present. This would be
discussed as a remaining issue. 
In contrast, 
we can construct the 1RSB theory irrespective of details of the model,
even non-mean-field model in principle, 
because our scheme does not rely on the microscopic details, or complexity.
Since the pure state in finite dimensions is difficult to formulate in a 
tractable manner, this complexity-independent formalism of 1RSB may be
useful to investigate nature of RSB in finite dimensions\cite{T2007}.
Because the replica method itself is originally independent of calculus
of spin variables, this theoretical flexibility would give another
perspective if RSB is formulated within macroscopic level. 
Therefore, we consider that the cavity method and our method are
complementary in order to understand the nature of SG. The correspondence
of their results in this model has a significance because 
they should provide the same result in the intersection of their 
validity range. 

Unfortunately, the validity of our 1RSB solution could not be
established within the scheme because of the lack of AT analysis.
Some 
AT analyses for finite-connectivity models are 
recently proposed in the previous works\cite{MR2003,OKN2008,KZ2008}.
They are to be resolved for our model  
and compared with each other in future study.
To confirm the validity of our scheme in the present work, equilibrium
MC simulations with the help of extended ensemble method have been performed for the model
with $K=2$ and $3$.  
It is shown that for $K=2$, the resulting thermodynamic quantities by
our scheme 
are in agreement with those obtained by MC simulation within statistical
error.
For $K=3$, 
assuming that the size dependence of the thermodynamic quantities is
expressed as 
a polynomial of $N^{-1}$, 
we have concluded that our 1RSB solution is 
also consistent with those extracted from the finite-size MC data. 
If we have the correction exponent $\omega$ a priori, 
we can promote the accuracy of our extrapolation. 
Optimization techniques for ground-state search would be a promising
approach for estimating the value of $\omega$ for $K=3$. 

As a by-product of the MC simulations, it is found that a coefficient of
the first finite-size-correction term is positive.
 Namely, the finite-size data reach their
thermodynamic value from above with increasing the system size. 
This suggests that fluctuations on the positive side of the
thermodynamic value is relevant for the finite-size corrections in these
models. On the other hand, 
the probability of large deviations
which can be calculated using the replica theory with $n>0$ 
is the negative side
for the free energy in the fully connected SK model\cite{PR2008}.
The replica theory with $n<0$ for the large deviations is required to
evaluate the finite-size correction. 

The key ingredient in our scheme for constructing the 1RSB solution is
the thermodynamic constraints as a necessary condition in the replica
theory. This is compared to the fact that the standard replica method
introduces RSB scheme through the symmetry of the saddle point. 
Another thermodynamic constraint, thermodynamic
homogeneity, has been discussed in Ref.~\onlinecite{JZ2005}.
One might stress the importance of such a thermodynamical approach which
leads to an universal framework irrespective of microscopic models. 
Actually, 
our scheme is rather general and quite simple.  
It only needs the function $\phi_{\rm RS}(n)$ which is 
constructed in the way of replica symmetric analysis. Thus, we can
avoid the arbitrariness to introduce breaking parameter in the replica
theory. 
One can find further applications in related statistical-mechanical 
systems in which the RS solution can be constructed. 

\begin{acknowledgments}
We would like to thank Y. Kabashima for helpful comments and
discussions, and for explaining the details of algorithm for solving the
 saddle-point equations\cite{KSNS2001}. We are also grateful to
 F. Krzakala for making his numerical data in Ref.~\cite{KZ2008}
 available to us. 
TN also gratefully acknowledges K. Mimura for the kind and
helpful lecture. He is strongly inspired by the lecture. 
This work was 
supported by the Grant-in-Aid for Scientific
Research on the Priority Area ``Deepening and Expansion of  Statistical
Mechanical Informatics'' (No. 1807004) by Ministry of Education,
 Culture, Sports, Science and Technology, Japan. 
\end{acknowledgments}

\appendix

\section{Algorithm for evaluating $\pi$ and $\hat{\pi}$}
\label{apppi}

\begin{figure}
\includegraphics[width=\figwidth]{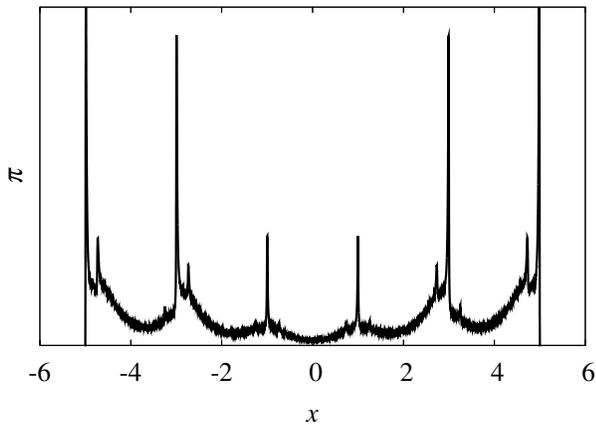}
\caption{A saddle-point function $\pi(x)$ for  $K=3, C=6, T=0.75$ and $n=0.713$.}
\label{funcpi}
\end{figure}

In this appendix, we explain details of the numerical method we used  to
solve the saddle-point equations (\ref{iter1}) and (\ref{iter2}). We use
an iteration method, introduced in Ref.~\onlinecite{KSNS2001}. The saddle-point
functions $\pi(x)$ and $\hat\pi(\hat{x})$ are approximated
by a large $M$ number of samples from $\pi$ and $\hat \pi$.
The algorithm for evaluating the function $\pi(x)$ and $\hat \pi(\hat{x})$ 
is as follows: 

\begin{enumerate}
\item Give an appropriate array $h_i(i=1,2,\cdots M)$ 
as an initial condition to $\pi$. 

\item Sample $K-1$ independent values of $\{x_k\}$ $(k=1,\cdots,K-1)$ from
$\pi(x)$ by generating a random integer $I$ uniformly distributed from 1
      to $M$ and setting $x_k =h_I$, 
and evaluate $\displaystyle \hat x = \frac{1}{\beta}{\rm atanh}
\left( \tanh\beta \prod_{k=1}^{K-1}\tanh(\beta x_k)\right)$. 

\item Put the sign chosen with probability 1/2 to $\hat{x}$ and get
      $\hat{h}_i=\hat{x}$, 
which corresponds to a sample of $\hat \pi$.

\item Repeat the steps 2 and 3 $M$ times and obtain the $M$ samples
       of $\hat \pi$, $\{\hat{h}_i\}$.

\item Sample $C-1$ independent values of $\{\hat x_\gamma\}$ $(\gamma=1,\cdots,
C-1)$ by a procedure similar to that of step 2
and evaluate $\displaystyle x=\sum_{\gamma=1}^{C-1}\hat
x_\gamma$. 

\item Accept $x$ obtained in step 5 with probability $\displaystyle
\frac{1}{2^{n(C-1)}} 
\left\{\prod_{\gamma=1}^{C-1}(1+\tanh(\beta \hat x_\gamma))+
\prod_{\gamma=1}^{C-1}( 1-\tanh(\beta \hat x_\gamma))\right\}^n$
and accumulate a new set of $\{{h}_i\}$ of $\pi$ till the number reaches $M$.

\item Return to 2.  
\end{enumerate}

We iterate the above procedures until convergence. 
The number of the samples is set typically as $M=10^6$ in our calculation. 
A typical form of $\pi(x)$ is displayed in Fig.~\ref{funcpi}. 


\end{document}